\documentclass[12pt]{article}
\usepackage{amsmath}
\usepackage{epsfig}
\begin{document}
\title{Nonlocality of the scalar quark condensate}
\author{M. G. Ryskin, E. G. Drukarev, V. A. Sadovnikova\\
{\em National Research Center "Kurchatov Institute"}\\
{\em B. P. Konstantinov Petersburg Nuclear Physics Institute}\\
{\em Gatchina, St. Petersburg}}
\date{June 7, 2014}
\maketitle

\begin{abstract}
We suggest a somewhat indirect method for estimation of the nonlocal scalar quark condensate. The approach is based on analysis of  the polarization operator of nucleon current in instanton medium.
\end{abstract}

In this Letter we demonstrate how calculation of the polarization
operator of the nucleon current in the instanton vacuum enables to find the value of the
nonlocal scalar quark condensate.

The nowadays belief is that the $\bar qq$ pairs which compose the vacuum condensate are created by instantons. The average size of instantons is $\rho=0.33\,{\rm fm}$, and they are separated by the distances of the order $R\approx1\,{\rm fm}$.
The theory of the small size instantons ($\rho \ll R$) was developed in \cite{1,2}, where the propagator of a light quark in the instanton medium $S(p)$ was found in an explicit form.
The scalar condensate is connected with the quark propagator by the relation
\begin{equation}
\langle 0|\bar q(0)q(0)|0\rangle=i\int\frac{d^4p}{(2\pi)^4}{\rm Tr} S(p).
\label{2}
\end{equation}

The coordinate $x$-distribution of quarks which compose the $\bar qq$ pair is described by the fermion zero mode wave function in the instanton field.
Thus one should investigate the $x$ dependence of the expectation value $\langle 0|\bar q(0)q(x)|0\rangle$. Calculation of the nonlocal condensate is a complicated task. At $|x^2| \ll \rho^2$ one can employ the Taylor expansion. However, for applications one needs the $x$ dependence at $|x^2| \geq \rho^2$. Here we suggest a none-straightforward way for determination of the function
\begin{equation}
f_q(x^2)=\langle 0|\bar q(0)q(x)|0\rangle.
\label{3}
\end{equation}
 We shall focus on the case $x^2 \sim (1\,{\rm GeV})^{-2}$.

Note that the product $\bar q(0)q(x)$ is not gauge invariant. This expression makes sense if we define $q(x)$ as the Taylor expansion near the point $x=0$, i.e.
\begin{equation}
q(x)=(1+x^{\mu}D_{\mu}+\frac{x^{\mu}x^{\nu}}{2}D_{\mu}D_{\nu}+...)q(0),
\label{3a}
\end{equation}
with $D^{\mu}$ standing for covariant derivatives.

The key idea is to analyze the proton polarization operator in the instanton vacuum. The polarization operator describes the time-space propagation of a system with the quantum numbers of the proton. It can be written as
\begin{equation}
\Pi(q^2)=i\int d^4xe^{i(q\cdot x)} \langle 0|T[j(x)\bar j(0)]|0 \rangle,
\label{3b}
\end{equation}
where the current $j(x)$ carries the quantum numbers of the proton, $q$ is the four-momentum of the system.
We employ the current
\begin{equation}
j(x)=(u^T_a(x)C\gamma_{\mu}u_b(x))\gamma_5 \gamma^{\mu}d_c(x) \varepsilon^{abc}.
\label{4} \end{equation}
suggested in \cite{3}.
The polarization operator contains the divergent terms originated by the behavior of the integrand at small values of $x$. They are the polynomials in $q^2$ and can be eliminated by the Borel transform.
We analyze the Borel-transformed operator ${\cal B}\Pi(q^2)=32\pi^4{\cal P}(M^2)$ at $M^2 \sim 1\,{\rm GeV}^2$. At $M^2\rho^2 \gg 1$ it can be expanded in powers of $1/M^2$, but this can not be done for $M^2\rho^2 \leq 1$.
We demonstrate, however, that the Borel transformed chirality flipping structure of polarization operator in the instanton vacuum
can be approximated by the sum of  three terms of the $1/M^2$ series.

We carry out calculations of the polarization operator under the standard assumption of the zero-mode domination.
This means that in the quark propagator
$ G(x,0)=\sum_n|\psi_n(x)\rangle\langle\bar\psi_n(0)|$
we treat the term with $n=0$ (zero mode) separately, while the sum of higher states (nonzero mode)
is approximated by the propagator of the free massless quark.
At the distances of the order
$x \sim 1\,{\rm GeV}^{-1}$ the quarks which compose the polarization operator can interact with only one instanton. Moreover, two $u$-quarks can not occupy the same zero mode while the contribution of $u$ and $d$ quarks, both in zero mode, vanishes in polarization operator with the current (\ref{4}) due to its spin structure. Thus we include the instanton effects only in one propagator, which can be written as
\begin{equation}
S_{ab}(p)=\frac{\hat p}{p^2}\delta_{ab}+\frac{im(p)}{p^2}\delta_{ab}.
\label{7}
\end{equation}
Here the second term on the right hand side approximates the zero mode contribution, while the first term is just the propagator of  free massless quark.
Thus the $u$ quarks in polarization operator are described by the free propagators, while the $d$ quark is described by the second term on the right hand side of Eq.~(\ref {7}).

On the other hand one can present the chirality flipping component of polarization operator as
\begin{equation}
\Pi(q^2)=\frac{2}{\pi^4}\int \frac{d^4x}{x^6}f(x^2)e^{iqx}\ .
\label{5}
\end{equation}
If the function $f(x^2)$ can be approximated as
\begin{equation}
f(x^2)=f(0)(1+c_1x^2+c_2x^4),
\label{6}
\end{equation}
in the region $x^2 \leq 1\,{\rm GeV}^{-2}$,  one can present the  coefficients $c_n$ in terms of the coefficients of expansion of the Borel transformed polarization operator.
Here $x^2$ is presented in psedoeucledian metric.
Note that the right hand side of Eq.~(\ref{6}) can not be treated as the lowest terms of the Taylor series for the function $f(x^2)$. The terms $x^{2n}$
with $n \geq 3$ of the Taylor series would provide the divergent terms caused by behavior at large $x$ of the integrands in the integrals on the right hand side of (\ref{5}). Such terms are not be eliminated by the Borel transform.

In the medium of the small size instantons the $u$ quarks and $d$ quarks of polarization operator are described by the first and second terms of the propagator (\ref{7}) correspondingly.
To obtain results in analytical form we parameterize the dynamical quark mass as
\begin{equation}
m(p)=\frac{{\cal A}}{(p^2+\eta^2)^3}.
\label{9}
\end{equation}
The power of denominator insures the proper behavior
$m(p) \sim p^{-6}$ at $p \rightarrow \infty$ \cite{1,2}.
Parameters ${\cal A}$ and $\eta^2$ are chosen to match the values of $m(0)$  and  of $\langle 0|\bar q(0)q(0)|0\rangle$ for $\rho=0.33\,{\rm fm}$ and $R=1\,{\rm fm}$, employed in \cite{1,2}. This provides $\eta^2=1.26\,{\rm GeV}^2$. The parameter $\eta^2$ does not depend on $R$, dropping with $\rho$ as $\rho^{-2}$.
Direct calculation of the polarization operator provides
\begin{equation}
{\cal P}(M^2)=2M^4a(M^2); \quad a(M^2)=a\cdot F(\frac{\eta^2}{M^2});\quad a=-(2\pi)^2\langle 0|\bar q(0)q(0)|0\rangle; \label{10}
\end{equation}
$$F(\beta)=\frac{2(1-e^{-\beta})}{3\beta}+\frac{1}{3}e^{-\beta}(1-\beta) +\frac{\beta^2}{3}E_1(\beta),$$
with
$$E_1(\beta)=\int_{\beta}^{\infty}dt\frac{e^{-t}}{t},$$
and $F \rightarrow 1$ at $M^2 \rightarrow \infty$.

Now we define
\begin{equation}
K(M^2)=\frac{a(M^2)}{a}
\label{80c}
\end{equation}
and try to find the function $K(M^2)$ as a power series in $1/M^2$
\begin{equation}
K(M^2)=1+\sum_{n=1}^{N}C_n/M^{2n}.
\label{80d}
\end{equation}
On the other hand we can write employing Eq.~(\ref{6})
\begin{equation}
{\cal P}(M^2)=2aM^4\left(1+\frac {8c_1}{M^2}+\frac{32c_2}{M^4}\right).
\label{11}
\end{equation}
Thus, if we manage to approximate the function $K(M^2)$ by Eq.~(\ref{80d}) with $N\leq 2$, we can identify
\begin{equation}
c_1=\frac{C_1}{8}; \quad c_2=\frac{C_2}{32}.
\label{12}
\end{equation}
We shall look for the solution in the interval $0.8\,{\rm GeV}^2 \leq M^2 \leq 1.4\,{\rm GeV}^2$ traditional for the sum rules analysis.

We find
\begin{equation}
C_1=-1.23\,{\rm GeV}^2; \quad C_2=0.54\,{\rm GeV}^4.
\label{90}
\end{equation}
The accuracy of the solution is illustrated by Fig.1.
Employing Eq.~(\ref{12}) we find $c_1=-0.16\,{\rm GeV}^2$ and $c_2=0.017\,{\rm GeV}^4$.
The nonlocality of the quark condensate can be described by the function
\begin{equation}
\kappa(x^2)=\frac{f(x^2)-f(0)}{f(0)}=c_1\cdot x^2+c_2\cdot x^4.
\label{91}
\end{equation}
Thus we obtain $\kappa=-0.14$ for $x^2=1\,{\rm GeV}^{-2}$.\\

Note in that much more complicated calculations of the function $f(x^2)$ in framework of the instanton liquid model
\cite{Shur} provided $\kappa \approx -0.1$ at $x^2=1\,{\rm GeV}^2$.

There were several moves to estimate the parameter $m_0$ defined as
\begin{equation}
m^2_0 \equiv \frac{\langle 0|\bar q\sigma_{\mu\nu}{\cal G}_{\mu\nu}q|0\rangle}{\langle 0|\bar q q|0\rangle};
\quad {\cal G}_{\mu\nu}=\frac{\alpha_s}{\pi}\sum_hG^h_{\mu\nu}\lambda^h /2,
\label{3dd}
\end{equation}
with $G^h_{\mu\nu}$ the tensor of the gluon field. This parameter may be used to characterize the quark condensate nonlocality since it is connected with second derivative at $x^2=0$ via the equation of motion. For the massless quark
\begin{equation}
\langle 0|\bar d(0)D^2d(0)|0\rangle=\frac{1}{2}\langle 0|\bar q\sigma_{\mu\nu}{\cal G}_{\mu\nu}q|0\rangle,
\label{4ddd}
\end{equation}
and thus $m^2_0$ determines the lowest term of the Taylor series for the function $f(x^2)$, i.e.
\begin{equation}
\kappa(x^2)=-\frac{m^2_0}{4}x^2+O(x^4).
\label{92}
\end{equation}
The lattice calculations \cite{Kremer} provided $m_0^2=1.1\,{\rm GeV}^2$.
The value of $m_0^2$ was evaluated also from the nucleon QCD sum rules analysis, where it was chosen to obtain
the best fit between the two sides of the sum rules. The estimation of \cite{I7} gave
$m_0^2=(0.8 \pm 0.4)\,{\rm GeV}^2$ \cite{I7}, while a smaller value $m_0^2=0.2\,{\rm GeV}^2$ was proposed in  \cite{1a}.
Note that the value of $m_0^2$ determines only the lowest order term of the Taylor expansion of the condensate near the point $x^2=0$, while in the sum rule analysis it actually imitates the whole effect of nonlocality of the scalar condensate. Therefore these estimations of the value of $m_0^2$ are consistent with our results.

We can consider a more general case when only a fraction $w_s\langle 0|\bar q(0)q(0)|0\rangle$ of the expectation value $\langle 0|\bar q(0)q(0)|0\rangle$ ($w_s<1$) is caused by the small size instantons while the origin of the rest part $(1-w_s)\langle 0|\bar q(0)q(0)|0\rangle$ is not clarified.
In this case
\begin{equation}
a(M^2)=a\cdot\left[1-w_s+w_sF(\frac{\eta^2}{M^2})\right]
\label{100}
\end{equation}
For example, in the case $w_s=0.65$
when the QCD sum rules reproduce the physical value of the nucleon mass \cite{5} we find  $C_1=-0.80\,{\rm GeV}^2$ and $C_2=0.35\,{\rm GeV}^4$. Thus $c_1= -0.10\,{\rm GeV}^2$ and $c_2= 0.011\,{\rm GeV}^4$. This provides $\kappa=-0.09$ for $x^2=1\,{\rm GeV}^{-2}$.

Thus our results are consistent with those obtained in other approaches.
They can be used in applications. For example, they can be employed in analysis of the QCD sum rules.

We acknowledge the partial support by the RFBR grant 12-02-00158.

{}
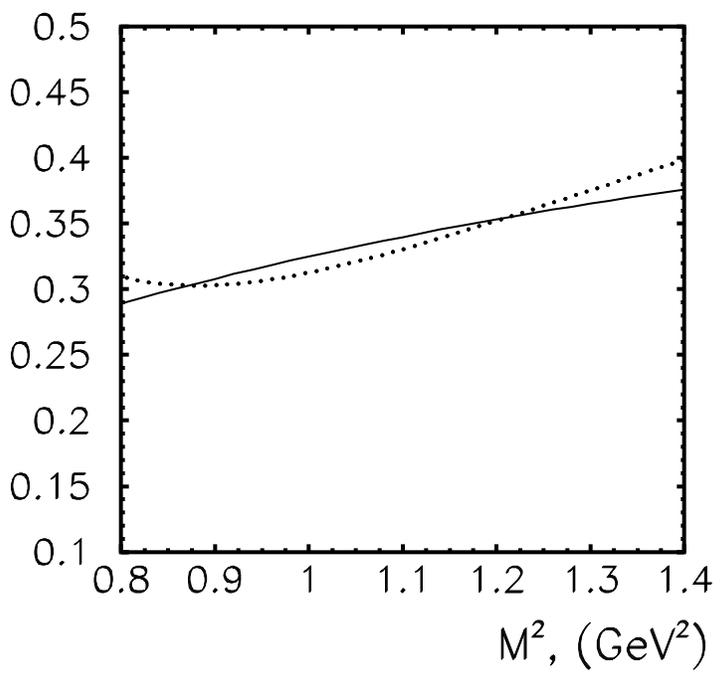
\end{document}